\documentclass[10pt,aps,twocolumn,superscriptaddress,longbibliography]{revtex4-2}

\usepackage[utf8]{inputenc}
\usepackage{enumerate}
\usepackage{braket}
\usepackage{upgreek}

\usepackage{amsmath}
\usepackage{amsthm}
\usepackage{amssymb}
\usepackage{float}
\usepackage{bbm}
\usepackage{amsfonts}
\usepackage{dsfont}
\usepackage{graphicx}
\theoremstyle{plain}
 
\usepackage{float}
\usepackage{hyperref}
\hypersetup{
pdftitle={Lattice surgery for near-term experimental logical qubit entanglement creation in planar architectures},
pdfauthor={Lukas Boedeker}
}
\usepackage[nameinlink,noabbrev]{cleveref}
\makeatletter
\AtBeginDocument{%
  \renewcommand{\selectlanguage}[1]{}%
}
\makeatother
\crefname{figure}{Fig.}{Figs.}
\Crefname{figure}{Fig.}{Figs.}

\crefname{table}{Table}{Tables}
\Crefname{table}{Table}{Tables}

\crefname{equation}{Eq.}{Eqs.}
\Crefname{equation}{Eq.}{Eqs.}

\crefname{section}{Sec.}{Secs.}
\Crefname{section}{Sec.}{Secs.}
\usepackage{fancyhdr}
\usepackage{xcite}
\usepackage{color}
\usepackage[dvipsnames]{xcolor}
\usepackage{MnSymbol}
\usepackage{bm}
\usepackage{bigints}
\usepackage{enumerate}

\usepackage{rotating}
\usepackage{comment}

\begin{document}

\title{Lattice surgery for near-term experimental logical qubit entanglement creation\\ in planar architectures}
\author{Lukas~B\"{o}deker}
\thanks{l.boedeker@fz-juelich.de}
\affiliation{Institute for Theoretical Nanoelectronics (PGI-2), Forschungszentrum J\"{u}lich, 52428 J\"{u}lich, Germany}
\affiliation{Institute for Quantum Information, RWTH Aachen University, 52056 Aachen, Germany} 
\author{Áron~Márton}
\affiliation{Institute for Theoretical Nanoelectronics (PGI-2), Forschungszentrum J\"{u}lich, 52428 J\"{u}lich, Germany}
\affiliation{Institute for Quantum Information, RWTH Aachen University, 52056 Aachen, Germany} 
\author{Luis~Colmenarez}
\affiliation{Institute for Theoretical Nanoelectronics (PGI-2), Forschungszentrum J\"{u}lich, 52428 J\"{u}lich, Germany}
\affiliation{Institute for Quantum Information, RWTH Aachen University, 52056 Aachen, Germany}
\author{Ilya~Besedin}
\affiliation{Department of Physics, ETH Zurich, 8093 Zurich, Switzerland}
\affiliation{ETH Zurich - PSI Quantum Computing Hub, Paul Scherrer Institute, 5232 Villigen, Switzerland}
\affiliation{Quantum Center, ETH Zurich, 8093 Zurich, Switzerland}
\author{Andreas~Wallraff}
\affiliation{Department of Physics, ETH Zurich, 8093 Zurich, Switzerland}
\affiliation{ETH Zurich - PSI Quantum Computing Hub, Paul Scherrer Institute, 5232 Villigen, Switzerland}
\affiliation{Quantum Center, ETH Zurich, 8093 Zurich, Switzerland}
\author{Markus~M\"{u}ller}
\affiliation{Institute for Theoretical Nanoelectronics (PGI-2), Forschungszentrum J\"{u}lich, 52428 J\"{u}lich, Germany}
\affiliation{Institute for Quantum Information, RWTH Aachen University, 52056 Aachen, Germany}

\date{\today}

\begin{abstract}
In the era of early fault-tolerant quantum computing, basic demonstrations of entanglement operations between a few logical qubits are at the frontier of recent developments in quantum computing. In this work, we describe in detail, at both the logical and physical qubit levels, a logical teleportation protocol between two surface code logical qubits based on lattice surgery. We address several aspects of the teleportation protocol pertinent to superconducting qubit architectures. We explore the modularity constraints in the number and location of stabilizer readouts and compare variants of the teleportation protocol in this regard. Additionally, we investigate potential performance improvements related to in-sequence decision logic and the optimal size of the interface region between two surface code patches on a superconducting chip.
Based on our simulations, we show possible near-term improvements in lattice surgery protocols that facilitate fault-tolerant quantum computing in superconducting circuit architectures.
\end{abstract}

\maketitle

\section{Introduction}

Quantum computers promise to process quantum information and perform certain computationally demanding tasks potentially faster than classical computers~\cite{sood_quantum_2024,dalzell_quantum_2025}, such as prime factorization \cite{shor_algorithms_1994}, and unstructured search \cite{grover_fast_1996}. On the path towards developing scalable fault-tolerant (FT) quantum computers, running logical subroutines such as logical gates \cite{Ryan_Anderson2024,hetenyi_creating_2024,Zhang2026,bluvstein_logical_2024}, preparation of non-Clifford states \cite{rosenfeld2025,sales_rodriguez_experimental_2025,daguerre_experimental_2025,pogorelov_experimental_2025}, error suppression after syndrome extraction and correction \cite{google_quantum_ai_and_collaborators_scaling_2025,google_quantum_ai_and_collaborators_suppressing_2023,postler_demonstration_2024,wang_demonstration_2026}, and execution of small quantum algorithms \cite{bluvstein_fault-tolerant_2026,Yamamoto2026,reichardt_fault-tolerant_2025} are important milestones for today’s near-term quantum devices. In particular, FT protocols require performing quantum error correction (QEC) to protect quantum information from noise and hardware imperfections~\cite{perlin_fault-tolerant_2026,wang2026,Besedin2026,rosenfeld2025,pogorelov_experimental_2025}. QEC codes realize encodings of single or multiple logical qubits into many physical qubits~\cite{gottesman_stabilizer_1997,terhal_quantum_2015}.
Quantum algorithms are then run at the level of logical qubits, while the physical qubits at a lower level are used for performing QEC and for implementing the logical gates. Therefore, successful QEC requires coherent control of many physical qubits at different levels, which poses a challenge to current quantum computing platforms. Although several quantum computing platforms are able to host dozens of physical qubits \cite{google_quantum_ai_and_collaborators_quantum_2025,rosenfeld2025,google_quantum_ai_and_collaborators_scaling_2025,Besedin2026,pogorelov_experimental_2025,postler_demonstration_2024,daguerre_experimental_2025,Ryan_Anderson2024,sales_rodriguez_experimental_2025,krinner_realizing_2022,wang2026}, and some even hundreds \cite{bluvstein_fault-tolerant_2026,bluvstein_logical_2024,hetenyi_creating_2024,reichardt_fault-tolerant_2025}, operating individual logical qubits and performing short error-corrected quantum algorithms on a few of them remains a challenging task.

Superconducting qubits are one of the most established systems for implementation of error correction codes. The straightforward approach to scaling superconducting qubits involves a 2D periodic arrangement where qubits are placed on a surface of a substrate, or possibly few stacked substrates~\cite{Foxen2018, Yost2020, Kosen2022}, where gates are available between neighboring qubits, with state-of-the-art systems having tens and few hundreds of physical qubits~\cite{Ye2023a, Gupta2024,google_quantum_ai_and_collaborators_scaling_2025,rosenfeld2025,google_quantum_ai_and_collaborators_quantum_2025,hetenyi_creating_2024,wang2026}. These properties make them naturally suitable for surface codes and lattice surgery operations, where only nearest-neighbor interactions are required. While the short-range coupling is a constraint that limits the class of codes that can be mapped efficiently onto this geometry, it also constrains the spread of errors, since most predominant error mechanisms are also short-range.

For this reason, topological codes such as the surface codes \cite{kitaev_quantum_1997,dennis_topological_2002} are currently among the primary QEC codes used in the pursuit of superconducting-circuit based FT quantum computers. Several demonstrations of quantum memories, i.e.~extending the coherence time of a logical state by successive rounds of error correction, have already been realized in single logical qubits \cite{andersen_repeated_2020, zhao_realization_2022,krinner_realizing_2022}, including better-than-break-even performance \cite{google_quantum_ai_and_collaborators_suppressing_2023,google_quantum_ai_and_collaborators_quantum_2025, google_quantum_ai_and_collaborators_scaling_2025}.
However, implementing a logical entangling operation between two logical qubits on planar architectures remains a difficult task, as transversal entangling operations cannot be used due to the inherent connectivity constraints.
Therefore, alternative schemes for creating entanglement between two logical qubits must be employed \cite{raussendorf_fault-tolerant_2007, raussendorf_topological_2007, fowler_surface_2012}, which often involve a large overhead in extra physical qubits and gate count. Lattice surgery \cite{horsman_surface_2012, fowler_low_2019} is today the most resource-efficient approach, as it utilizes a few auxiliary physical qubits between two surface code patches to measure a joint operator of two logical qubits~\cite{fowler_low_2019}.
Several proposals exist for performing lattice surgery in superconducting qubit arrays \cite{ueno_qulatis_2022, tan_sat_2024, leblond_realistic_2023} and first experiments have demonstrated lattice surgery in the context of repetition codes, color codes and the surface code~\cite{Besedin2026,google_quantum_ai_and_collaborators_scaling_2025,wang2026}.
The lattice-surgery technique is vital, not only for superconducting-circuit-based qubits, but also for other experimental platforms. Demonstrations of its effectiveness have been successfully conducted in trapped ion systems~\cite{erhard_entangling_2021,Ryan_Anderson2024}. Aligned with first demonstrations~\cite{Besedin2026,google_quantum_ai_and_collaborators_scaling_2025,wang2026}, obtaining a modularly extensible and scalable superconducting quantum processor capable of performing lattice surgery represents a significant step toward universal FT quantum computation. This is because the full logical Clifford group can be implemented fault-tolerantly using lattice surgery \cite{chatterjee_lattice_2024}, and logical non-Clifford operations can be achieved through magic state distillation or cultivation and injection \cite{gidney_magic_2024,gavriel_transversal_2023, fowler_low_2019, fowler_surface_2012,google_quantum_ai_and_collaborators_scaling_2025, rosenfeld2025}.

One of the fundamental demonstrations of entanglement between two qubits is to teleport a qubit using a logical Bell pair \cite{bennett_teleporting_1993}. Hence, the entanglement is not only created but also \emph{used} in the teleportation protocol. In this work, we study two-qubit fault-tolerant surface code-based teleportation protocols designed for superconducting qubit chips. 
Specifically, we focus on projecting the performance of such protocols on what we call the \emph{surface-41} chip (see~\cref{fig:surface_41_layout}), which can be viewed as two surface-17 patches \cite{krinner_realizing_2022, zhao_realization_2022} with a column of three physical data qubits in between, or as a rectangular $3\times7$ rotated surface code.
We present stabilizer simulations of these protocols on the surface-41 chip, using experimentally informed error rates~\cite{krinner_realizing_2022}. Moreover, based on the stabilizer simulations, we discuss the feasibility of the proposed protocols given current and future metrics of superconducting devices and explore directions for hardware improvements. 
Furthermore, we theoretically study broader aspects of the protocols, including optimization of the number of syndrome extraction measurements by means of dynamically adapted circuits, i.e.~by allowing that stabilizer measurements are conditioned on previous measurement outcomes. Based on numerical simulations, we estimate the parameter regimes in which such in-sequence decision logic is expected to lead to a higher fidelity of the logical state teleportation protocol. Then,  we explore the possible benefits in increasing the number of physical qubits between the two surface code patches. Thereby we explore the teleportation protocol for arbitrary distances and different sizes of the interface region. We find that when the protocol operates in the sub-threshold regime, there is no scenario in which adding extra qubits in the interface region increases the teleportation fidelity. Therefore, the optimal choice is to minimize the size of the interface region between two surface code patches.

The manuscript is organized as follows: 
We start in~\cref{sec:lattice_surgery}, by introducing the setting for lattice surgery applied to the rotated surface code, focusing on a superconducting qubit architecture.
To this end, we introduce the overall protocol of logical state teleportation, as well as an experimentally informed noise model which we use to estimate the logical performance. In the remainder of~\cref{sec:lattice_surgery}, we discuss possible alterations to the standard lattice surgery protocol in terms of its average circuit depth, still ensuring fault tolerance while improving logical fidelity. Lastly, in~\cref{sec:high_distance}, we investigate the scalability of logical state teleportation and the influence of the size of the merging region on logical performance. Overall, our results provide design guidelines for lattice surgery-based logical operations in near-and mid-term experimental devices.

\section{Low overhead lattice surgery}\label{sec:lattice_surgery}

In this section, we explain in detail the protocols proposed for two-qubit logical teleportation experiments based on lattice surgery. 
We start by outlining basic concepts and the protocols for lattice-surgery-based teleportation between two logical qubits. Based on this, we study how lattice surgery protocols can be optimized while maintaining fault-tolerance and modularity of the operation. 
We further discuss considerations specific to superconducting transmon-qubit platforms leading to a realistic assessment of projected experimental teleportation fidelities.

\subsection{Context and Theoretical Background}
\label{subsec:theory}

\begin{figure}
  \begin{center}
    \includegraphics[width=1\columnwidth]{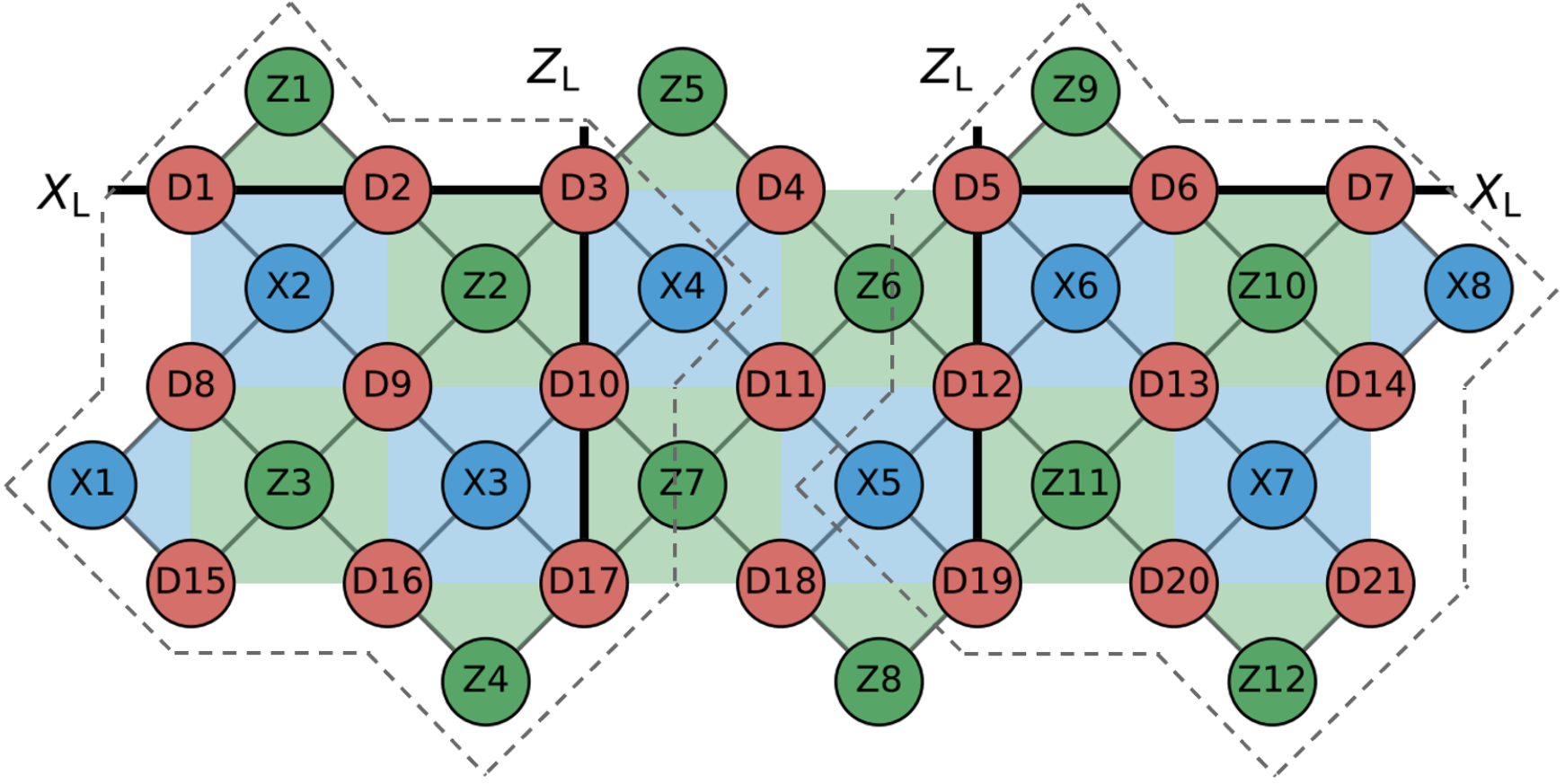}
  \end{center}
  \caption{Surface-41 layout. Blue (green) plaquettes denote $X$ ($Z$) stabilizers. Two independent surface-17 logical qubits, located on the right and left parts of the qubit layout, are enclosed by dashed lines. For the teleportation protocol, the right surface code patch is considered the source and the left is considered the target qubit. Data qubits are represented by red circles, including the three central qubits used during merging. Ancilla qubits are shown as blue and green circles, indicating the type of stabilizer they measure. Grey lines represent all physical connections between qubits.}
  \label{fig:surface_41_layout}
\end{figure}

\emph{The rotated surface code---} 
To provide a near-term perspective on the experimental realization of logical entanglement between planar error-correcting code patches, we focus on small instances of the rotated surface code~\cite{Bombin2007}.
The stabilizer generators of the surface code, which are measured to correct errors, are defined in terms of plaquette operators, namely weight-four plaquettes in the bulk of the lattice and weight-two plaquettes at the boundaries. These stabilizers are either of $X$ or $Z$ type and are arranged alternatingly in rows and columns of the underlying lattice structure. 
The logical $X$ and $Z$ operators of the surface code have minimal representations as two orthogonal boundary strings, each with minimal qubit support equal to the length of the corresponding boundary. The weight of the minimal representation of either of the logical operators is called the distance $d$. Under FT operation, the code can correct up to $(d-1)/2$ errors. 
In the next sections we will mostly focus on the $d=3$ instance, known as the Surface-17 code \cite{tomita_low-distance_2014,krinner_realizing_2022,google_quantum_ai_and_collaborators_quantum_2025}. Two realizations of this smallest error-correcting surface code are shown in~\cref{fig:surface_41_layout}, highlighted by dashed lines. The figure also illustrates the layout of a $3\times7$ rotated surface code.

\textit{Logical teleportation ---} In order to create entanglement between logical qubits, we perform a joint logical parity measurement of the Pauli operator $P_L$ on the (first) source and the potentially different Pauli operator $P'_L$ on the (second) target logical qubit. We omit the tensor product and denote the joint logical parity operator of the encoded quantum states as $P_L P'_L$, where subscript $L$ indicates that we consider objects on the logical level.

\begin{figure*}[t]
    \centering
    \includegraphics[width=\textwidth]{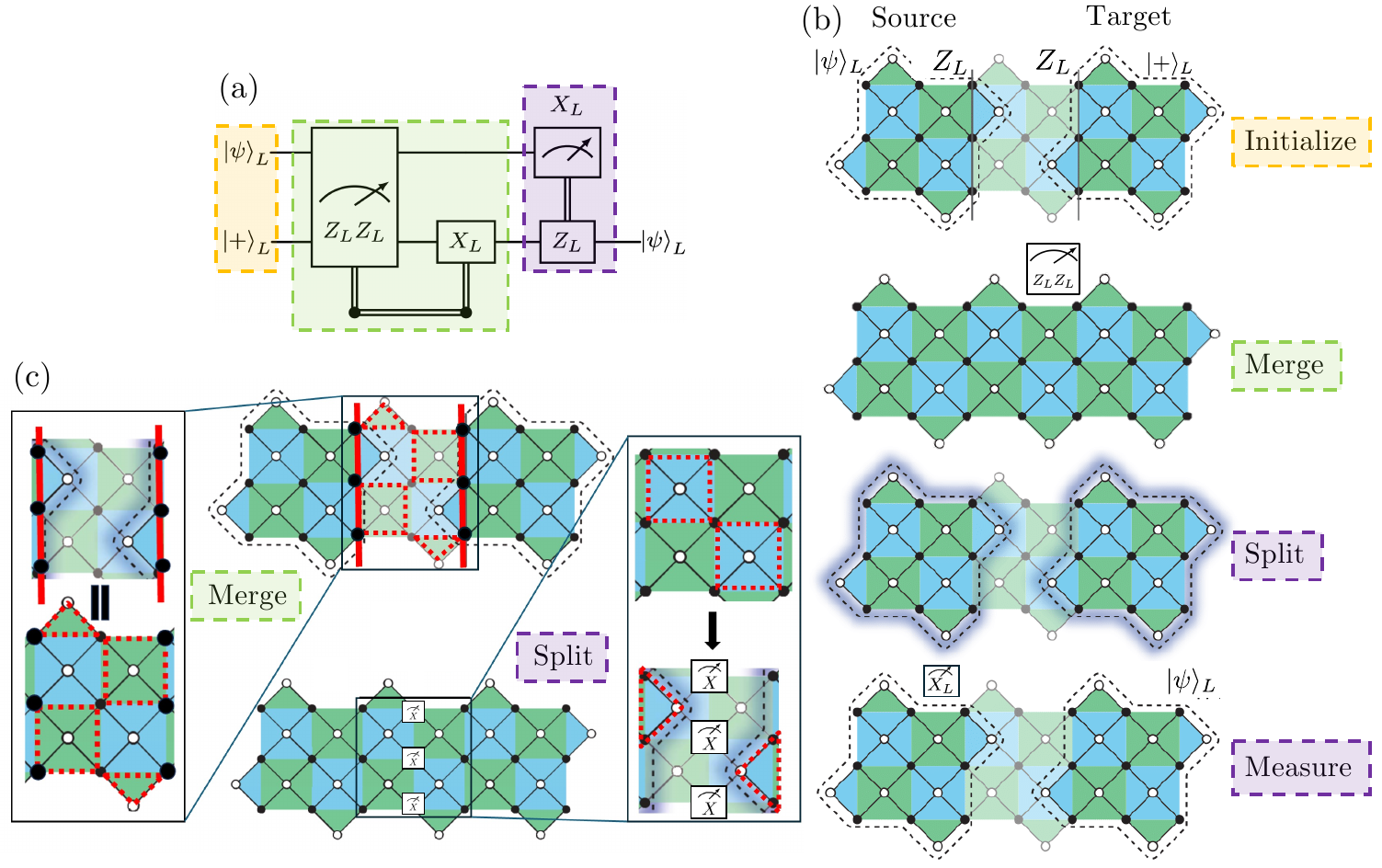}
    \caption{(a) High-level state teleportation scheme. Color blocks denote each step in the logical surface code circuit. (b) Sketch of quantum teleportation of two surface code patches via lattice surgery. 
    The initialization of the logical state (yellow) $|\psi\rangle_L$ on the left (source) logical qubit and the state $|+\rangle_L$ on the right (target) logical qubit. The joint measurement $Z_L Z_L$ is executed by “merging” (green) of the two individual surface codes. During the merging we consider the newly initialized $Z$-type stabilizers and the extended $X$-type stabilizers as the middle-part of the joint code. The final teleportation step shows the “split” step (violet) in which the individual logical qubits are recovered by measuring the physical qubits along the interface. Then the $X_L$ operator on the left qubit is measured (violet). The state $|\psi\rangle_L$ is teleported onto the right surface-17 patch. Any logical Pauli corrections based on the logical measurement outcomes can be performed in post-processing. (c) Illustration of merging and splitting operation during the lattice surgery. The respective zooms show the equivalence of the $Z_LZ_L$ to the product of the middle $Z$ stabilizer and the reduction of the middle $X$ stabilizers due to the measurement of the middle data qubits in the $X$ basis.}
   \label{fig:surgery_protocol}
\end{figure*}

In the following, we restrict the discussion to the logical $Z_L Z_L$ parity measurement where the surface code patches are oriented parallelly along the boundary where the $Z$-type logical operator has support. Orienting the patches along the $X$-type logical operators allows to measure $X_L X_L$ in a similar way~\cite{horsman_surface_2012,litinski_game_2019}.
The measurement of the logical parity between two code patches corresponds to measuring a weight-$2d$ operator, where $d$ is the distance of the codes. For performing lattice surgery, this large-weight operator is decomposed into a product of commuting operators. This set of lower-weight operators can be measured simultaneously with the help of local ancillary qubits. In the $d=3$ case, shown in~\cref{fig:surface_41_layout}, such ancilla qubits are denoted as $Z_5, Z_{6}, Z_{7}, Z_8$.
By measuring this set, the desired logical parity can be reconstructed from the product of the measurement outcomes. 
The lower-weight operators have support on the boundary of the surface code patches as well as on additional physical qubits located between the two logical qubits. In the $d=3$ case in~\cref{fig:surface_41_layout}, those additional qubits are denoted as $D_4, D_{11}, D_{18}$.
For the surface code, the operators that constitute the $Z_L Z_L$ parity can be seen as $Z$ stabilizers at the interface of the two codes. This decomposition is exemplified in~\cref{fig:surgery_protocol} (c) in the merging step. These stabilizers do not belong to either of the two codes. Rather, they are the stabilizers of the natural continuation between the two codes, as if one were to merge the two surface code patches into a single code. This procedure is illustrated in~\cref{fig:surgery_protocol}~(b).

We will use the tool of lattice surgery to measure the $Z_LZ_L$ parity and perform a deterministic teleportation of a logical state $\ket{\psi}_L$ from a logical source qubit to a logical target qubit, as depicted in~\cref{fig:surgery_protocol}~(a). This two-qubit teleportation circuit is identical for physical and logical qubits. For an arbitrary logical source state $\ket{\psi}=\alpha\ket{0}+\beta\ket{1}$ on the source qubit and $\ket{+}=2^{-1/2}(\ket{0}+\ket{1})$ on the target qubit, the teleportation procedure can be described with two measurement steps followed by conditional Pauli corrections, see~\cref{subsec:surgery_protocol} for the circuit. In the following explicit expression, the first state represents the source and second state the target qubit:
\begin{align}
\ket{\psi} \otimes \ket{+}  = & \dfrac{\alpha}{\sqrt{2}}\ket{00}+\dfrac{\alpha}{\sqrt{2}}\ket{01}+\dfrac{\beta}{\sqrt{2}}\ket{10}+\dfrac{\beta}{\sqrt{2}}\ket{11}\\
&\overset{ZZ}{\longrightarrow}\mathds{1}\otimes (X)^{m_{ZZ}}\left(\alpha\ket{00}+\beta\ket{11}\right)\\
&\overset{X\mathds{1}}{\longrightarrow} (X)^{m_{ZZ}}(Z)^{m_{X}}\left(\alpha\ket{0}+\beta\ket{1}\right),
\end{align}
where $m_{ZZ},m_{X}\in\{0,1\}$ are the measurement outcomes of the joint $ZZ$ parity and the $X$ parity of the source qubit, respectively, in the sense that outcomes $\{1,-1\}$ correspond to $\{0,1\}$. The operations $(X)^{m_{ZZ}}$ $(Z)^{m_{X}}$ are conditional logical Pauli corrections which are applied if and only if the respective logical measurement outcomes are 1.

The lattice surgery for conducting the logical teleportation is performed in four steps, which are illustrated for the $d=3$ surface-17 code in~\cref{fig:surgery_protocol} (b). It involves stabilizer measurements and the measurement of the auxiliary physical data qubits between the two logical qubits.

\begin{enumerate}[(i)]
\item Initialize (yellow in~\cref{fig:surgery_protocol}): We choose to initialize the logical qubits by projective stabilizer measurements. To this end, the data qubits on the target logical qubit are always prepared as $|+\rangle^{\otimes d^2}$ before projecting on the target $Z$ stabilizers. The data qubits of the target logical qubit correspond to the qubits $\{D_5, D_6, D_7,D_{12}, D_{13}, D_{14},D_{19}, D_{20},D_{21}\}$ for distance $d=3$ in~\cref{fig:surface_41_layout}. The initial state of the source logical qubit depends on the cardinal state that shall be teleported, i.e. we prepare it in  $\{\ket{0}^{\otimes d^2},\ket{1}^{\otimes d^2},\ket{+}^{\otimes d^2},\ket{-}^{\otimes d^2}\}$ for teleporting $|\psi\rangle_L\in\{\ket{0}_L,\ket{1}_L,\ket{+}_L,\ket{-}_L\}$ respectively. These data qubits of the source logical qubit patch are given by qubits $\{D_1, D_2, D_3, D_8,D_{9}, D_{10},D_{15},D_{16},D_{17}\}$ for distance $d=3$ in~\cref{fig:surface_41_layout}. 
The $d$ intermediate data qubits between the two code patches are initialized in $\ket{+}^{\otimes d}$. For the example of distance $d=3$ this corresponds to the qubits $\{D_4,D_{11},D_{18}\}$ in~\cref{fig:surface_41_layout}.
\item Merge: The $(d+1)$ $Z$ stabilizers at the interface of both patches are measured.  
The $Z_{L} Z_{L}$ operator measurement is then obtained as the product of the measurement outcomes of the $Z$ plaquettes in the interface of the two logical qubits, see~\cref{fig:surgery_protocol}(c). 
In the $d=3$ example in~\cref{fig:surface_41_layout}, the $Z$ stabilizers are measured using the ancilla qubits $\{Z_5, Z_6, Z_7, Z_8\}$. Since $Z_L Z_L$ has support on qubits $\{D_3,D_{10},D_{17},D_5,D_{12},D_{19}\}$, the product of the four $Z$ stabilizers equals the value of the $Z_L Z_L$. 

The middle weight-two $X$ stabilizers are extended by measuring them together with the intermediate data qubits which were initialized as $\ket{+}^{\otimes d}$. Thus, the previous stabilizer parities are preserved and the operators are extended to weight-four stabilizers (see~\cref{fig:surgery_protocol}c ).
For the $d=3$ example in~\cref{fig:surface_41_layout}, this concerns the $X$ stabilizer with support on $\{D_3,D_{10}\}$ that is extended to $\{D_4,D_{11}\}$ and the $X$ stabilizer with support on $\{D_{12},D_{19}\}$ that is extended to $\{D_{11},D_{18}\}$.

\item Split: The $d$ data qubits aligning vertically in the middle of the interface are measured in the $X$ basis (depicted in~\cref{fig:surgery_protocol} (c) in the split inset). This randomizes the inner $Z$ plaquette operators, as they do not commute with the measured operators. As a consequence, the code is reduced to two separated patches, see~\cref{fig:surgery_protocol}(b). 

\item Measure: All physical qubits of the logical source qubit are measured projectively in the $X$ basis to determine its $X_{L}$ parity as indicated in~\cref{fig:surgery_protocol} (b).
\end{enumerate}

All conditional logical operations on the target qubit, based on the measurement outcomes of the $X_L$ parity of the source logical qubit and the joint $Z_L Z_L$ parity (see~\cref{fig:surgery_protocol}), can be performed virtually or in post-processing~\cite{terhal_quantum_2015}.

\subsection{Fault-tolerance-preserving overhead reduction}
\label{subsec:overhead}
\label{subsec:surgery_protocol}
\label{subsec:FT_modularity}

In this section, we address the so-far omitted general considerations on the fault tolerance (FT) of lattice surgery-based protocols to then discuss concrete protocol variants. We will not resort to rigorous definitions for which we refer to~\cite{gottesman2010introduction,chamberland_flag_2018} and focus on the minimally correctable fault-weight instead.
As discussed before, the logical parity measurement that is conducted in the merging causes a conditional logical Pauli correction.
It is therefore essential to error-correct these measurement outcomes. For distance $d=3$, it must be avoided that, for instance, a single measurement error flips the logical parity measurement and therefore incurs a logical error. 
In a larger algorithm, with multiple logical qubits surgery operations and potentially non-Clifford operations, the logical corrections of the surgery operations must be determined and applied for every entangled block of logical qubits before the application of a layer of logical non-Clifford gates to maintain the prospect of quantum advantage~\cite{litinski_game_2019,sorathia_quantum_2025}. This defines the window in which the logical parities need to be measured with sufficient confidence.

In general, to determine a logical correction fault-tolerantly, i.e., in a way that ensures that errors occurring during the operation do not propagate into uncorrectable logical errors in a distance-$d$ code, one needs to be able to correct at least $\left\lfloor (d-1)/2 \right\rfloor$ faults~\cite{gottesman2010introduction}. Since single measurements can be faulty, this implies that one needs to measure the joint logical operator $d$ times.  
In other words, in a lattice surgery operation, the comprising stabilizers located at the interface of the two logical qubits must be measured $d$ times. The same principle applies in a memory experiment, where, to correct $\left\lfloor (d-1)/2 \right\rfloor$ faults, one needs to execute $d$ rounds of full stabilizer measurements~\cite{Fowler2012}. Note, however, that the according circuits are themselves noisy. 
The number of stabilizer readouts can be reduced, for instance to a minimum of $(d-1)/2$ when the execution of stabilizer measurements is conditioned on previous measurement outcomes in the merging operation of the lattice surgery protocol. The potential of such in-sequence decision logic and its potential benefit on the logical performance is discussed in more detail in~\cref{subsec:Adaptive_stab_meas}.
For the teleportation protocol, consequently, it is sufficient to execute the syndrome readouts needed for error-correcting the $Z_L Z_L$ parity and $X_L$ on the source logical qubit to ensure the fault-tolerance of the teleportation protocol, see~\cref{fig:surgery_protocol}. Therefore, there is some flexibility and room for circuit optimization in the correcting intermediate parts of the protocol, such as logical qubit initialization, as long as the minimum requirements to ensure fault-tolerance in correcting $Z_L Z_L$ and $X_L$ are satisfied.

\textit{Modularity---} One central aspect of the system layout depicted in~\cref{fig:surgery_protocol} (b) is its architectural modularity, i.e.~the fact that both logical qubits can be operated independently of each other. This includes the application of gates on physical qubits as well as the measurement of stabilizers. These capabilities, combined with measurement and initialization of individual qubits are sufficient to obtain a universal, error corrected quantum computer based on lattice surgery when the number of independent modules is scaled up~\cite{horsman_surface_2012,litinski_game_2019}. 
Thus, modularity of the logical qubits in the outlined teleportation protocol implies that stabilizers measurements can be conducted in parallel on both logical qubits, thereby allowing parallel detection and removal of errors on each logical qubit.
The lattice surgery operation can consequently be thought of as a logical gadget to be applied to neighboring independent logical qubits also in a bigger logical processor. The ability to error-correct not only the entire processor globally but also each logical operation is another expression of modularity~\cite{cain_correlated_2024}.

Following this idea for the surgery protocol, one can explore the question of modularity in terms of the scheduling of stabilizer measurements. Generally, a FT correction can be determined given the syndrome of $\mathcal{O}(d)$ full stabilizer measurements. 
The repeated measurement of all stabilizers that ensures FT error correction is referred to as a QEC round. In other words, during a QEC round, the necessary information for reliably identifying and removing errors is extracted from stabilizer measurements. For distance $d=3$, a QEC round consists of two rounds of measuring the $X-$ and $Z-$type stabilizers.

In the following, we will compare a setting for $d=3$ where in every stage of the lattice surgery a QEC round is conducted such that every step of the protocol can be corrected independently. Another setting is where the QEC round is distributed along different steps of the protocol such that the whole teleportation protocol remains FT. 
The former we call \textit{fully modular}, due to the ability to correct each part of the teleportation (initialization, merging, and readout) independently, and the latter we call \textit{depleted} due to the ability to correct only the whole teleportation protocol.
The core idea is that certain stabilizer measurements are not necessary for correcting the entire teleportation protocol but allow for FT corrections in individual logical blocks of the teleportation circuit.

\begin{figure*}
    \begin{center}
    \includegraphics[width=0.97\textwidth]{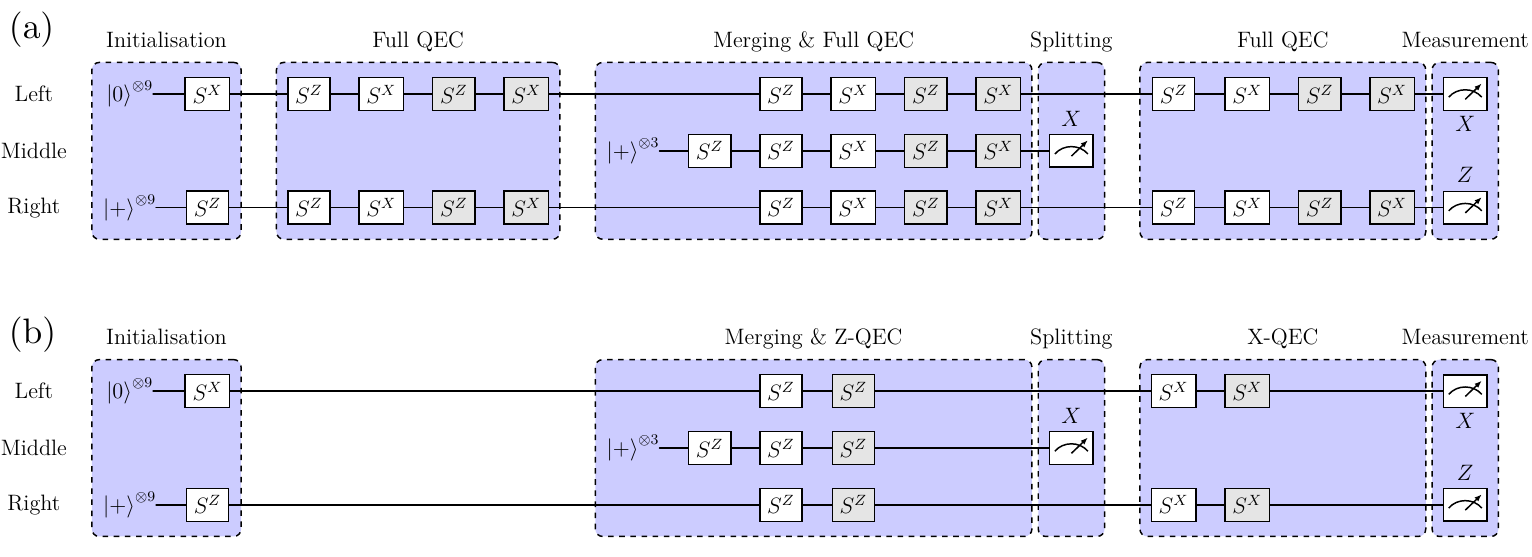}
  \end{center}
  \caption{Schematic comparison of the used lattice-surgery protocols. The boxes $S^X$ and $S^Z$ indicate stabilizer measurements on the respective parts of the chip. (a) \textit{Fully modular} teleportation scheme for the $\ket{0}_L$ state of the surface-17 code. The teleportation of $\ket{+}_L$ is analogous for initializing the left data qubits in $\ket{+}^{\otimes 9}$. All $X$ or $Z$ stabilizers (denoted by the boxes $S^X$ and $S^Z$) left (source) and right (target) surface code patch are measured in parallel. This includes the middle region while the patches are merged. The gray background indicates that the according stabilizer measurements can be made conditional on the previous outcomes.
  The ancilla measurement readouts are not shown; they are executed in parallel to other operations whenever possible to avoid idling times~\cite{krinner_realizing_2022}. Otherwise, they take place between time blocks. For example, while the $X$ stabilizer ancillas of the left logical qubit are being measured in the initialization step, the circuitry for the next $Z$ stabilizer measurement in the QEC round is executed already, see~\cref{fig:full_circuit_a,fig:full_circuit_b} in Appendix~\ref{sec:circuits}.  The measurements indicated at the end of each line denote measurements of the data qubits.
  (b)\textit{Depleted} teleportation scheme for the $\ket{0}_L$ state of the surface-17 code. The teleportation of $\ket{+}_L$, uses the same gate sequence as for initializing the left data qubits in $\ket{+}^{\otimes 9}$. The gray background indicates the according stabilizer measurements that can be made conditional on the previous outcomes. Immediately after each gate sequence for mapping a stabilizer parity to an ancillary qubit it is measured, if possible, in parallel with the subsequent gate sequence. The circuit details, gate scheduling and other lower-level details of the protocols at the physical qubit level are shown in Appendix~\ref{sec:circuits}.}
  \label{fig:combined_scheme}
\end{figure*}

\textit{Modularity and depletion ---}We contrast the two protocols shown in~\cref{fig:combined_scheme}, where the teleportation of a specific state is shown schematically in terms of when and which stabilizers are measured. For instance, in order to be able to correct for errors in the initialization phase in a self-contained (modular) way, one needs to perform respectively one independent QEC round on either of the logical qubits which entails two full rounds of measuring all stabilizers.
The scheme~\cref{fig:combined_scheme}~(a), that we call the \textit{fully modular} one, in each stage of the teleportation protocol, a full QEC round, is conducted such that one could determine independently FT corrections for the initialization, the merging, and the splitting. On the other hand, for the scheme in~\cref{fig:combined_scheme}~(b), that we call the \textit{depleted} one, we remove all stabilizer readouts that are not strictly necessary to guarantee FT of the entire teleportation, while still obtaining the necessary information to determine corrections for both the $X$ and the $Z$ basis.
 
Further, it must be stressed that QEC rounds after initialization are not necessary to preserve FT for the overall protocol, but only for enabling a FT correction decision at this initialization stage. 
This is fundamentally different with respect to the QEC round after the merging operation. 
Each of the middle $Z-$ stabilizer values must be determined such that no single fault inverts the true logical parity outcome upon which logical Pauli corrections are applied as laid out before. To ensure FT, the $Z$ stabilizers of the merged code must be remeasured after the projective measurement at the merging stage, as shown in~\cref{fig:combined_scheme}~(a) and~\cref{fig:combined_scheme}~(b).
Lastly, after merging, the two surface code patches are split by measuring the middle physical data qubits in the $X$ basis, yielding a random outcome. This randomizes the middle $Z$ stabilizers and reduces the support of the middle $X$ stabilizers to the original weight-two boundary stabilizers of the separate surface code patches, see~\cref{fig:surgery_protocol}. It must be taken into account that the values of these reduced stabilizers are given by the product of the old weight-four stabilizers and the measurement outcomes of the respective data qubit measurement outcomes, in order to maintain FT.

After splitting, the data qubits of the source logical qubit are measured transversally in the $X$ basis. From this measurement, the logical $X_L$ parity can be determined by reconstructing stabilizers from data qubit measurements that can be decoded together with previous stabilizer readouts to obtain corrected data qubit values with a well-defined logical parity.
Based on the derived logical $X_L$ parity, a logical correction operation ($Z_L$) is applied on the target logical qubit. Here, due to the transversal measurement, no single fault leading to data qubit errors or measurement errors can break fault-tolerance, even if no $X$ type syndrome measurement had been performed before.
To determine the teleportation success, the data qubits of the logical target qubit are also measured transversally in the respective basis to determine the logical parity value of the teleported cardinal state, upon a round of error correction via reconstructing the syndrome from data qubit measurements.
It must be noted that this depletion of the number stabilizer measurements during the merging is not scalable to larger distances as outlined in~\cref{sec:scala_deplt}. The reason is that the number or $Z$-type stabilizer measurements in the scheme increases with $d$ leading to an accumulation of $Z$-type errors on the data qubits and hence a vanishing threshold for correcting phase flip errors. For $d=3$, however, we still find an improvement due to the depletion as will be shown in~\cref{subsec:transmon}.

\subsection{Experimentally motivated noise modelling and benchmarking}
\label{subsec:transmon}

We base our simulations on a superconducting transmon qubit platform compatible with recent surface-code experiments~\cite{krinner_realizing_2022}. The purpose of outlining this architecture is not to prescribe a specific hardware implementation, but to define a physically motivated mapping from abstract circuits to a native gate set, together with realistic time scales and error parameters that enter the noise model. In particular, the chosen architecture determines the available gate primitives, their durations, and the dominant error mechanisms, which together define the circuit-level noise model used in our simulations.
We assume frequency-tunable transmon qubits with fixed nearest-neighbor coupling and a native gate set consisting of single-qubit rotations and CZ two-qubit gates. Gate implementations follow standard realizations in this platform, such as flux-controlled CZ interactions and microwave-driven single-qubit rotations~\cite{Negirneac2021,Lazar2022}. 

Rather than modeling the detailed pulse-level dynamics, we abstract these operations to their effective durations and average error rates. Specifically, we use representative gate times of 40~ns for single-qubit gates, $\sim$100~ns for two-qubit gates, and 400~ns for measurement, consistent with recent experiments (see Table~\ref{tab:Duration_times} in Appendix~\ref{sec:noise_modeling}). These time scales directly determine the decoherence-induced error during both gate execution and idle periods.
In the experiment, readout is performed as a dispersive measurement via resonators with frequency multiplexing, typically yields a finite assignment error probability.
In summary, at the circuit level, we work with a gate set composed of single-qubit $Y$ rotations and CZ gates, supplemented by noisy state preparation and measurement operations (see Appendix~\ref{sec:circuits} for details).

The physical noise processes of this architecture are captured using an incoherent circuit-level Pauli noise model. This model approximates the dominant error mechanisms—gate imperfections, measurement and initialization errors, and decoherence during idle times. This level of abstraction is sufficient for our purposes, as the primary goal of this work is to compare logical protocols and study their scaling behavior, rather than to reproduce device-specific microscopic noise features.

Concretely, the noise model includes: (i) stochastic errors in state preparation and measurement, (ii) depolarizing noise following single- and two-qubit gates, as well as (iii) decoherence and spontaneous de-excitation during idle periods parameterized by $T_1$ and $T_2$. The idling noise is incorporated through effective Pauli error channels whose probabilities depend explicitly on the duration of each idle segment. As a result, the hardware-dependent timing of circuit elements (Table~\ref{tab:Duration_times} in Appendix~\ref{sec:noise_modeling}) directly translates into idling error rates.
The base error parameters are taken from experimental results on a superconducting surface-code device~\cite{krinner_realizing_2022}, summarized in Appendix~\ref{sec:noise_modeling} Table~\ref{tab:ref_error_rates}. We do not consider other types of noise such as leakage or crosstalk explicitly. In view of the experimental reference~\cite{krinner_realizing_2022}, this can be justified as the experimental outcome can still be reproduced in simulations while making these simplifications.
Further, the measurement of leaked qubits can be post-selected upon such that undetected leakage is incorporated in the used noise parameters. This however does omit temporally correlated errors that can be caused by persistent leakage~\cite{Remm2026}. Crosstalk, which we understand as spatially correlated noise, is in fact negligible in the experimental context we consider~\cite{krinner_realizing_2022}.

\textit{Benchmarking ---} 
The noise reference values represent current hardware performance of the QuDev lab at ETH Zurich and define a reference operating point ($\lambda=1$) \cite{krinner_realizing_2022}. Lower error rates have already been demonstrated~\cite{google_quantum_ai_and_collaborators_scaling_2025}, but are still on the same order of magnitude and we argue that the structure of the modeled noise is similar compared to other implementations. To study such prospective improvements, we uniformly scale all error rates by a factor $\lambda \in (0,1]$, which allows us to estimate the impact of future hardware advances on logical error rates. This approach provides a controlled and interpretable connection between physical error rates and the performance of the logical protocols considered here.

\begin{figure}
    \centering
    \includegraphics[width=1\columnwidth]{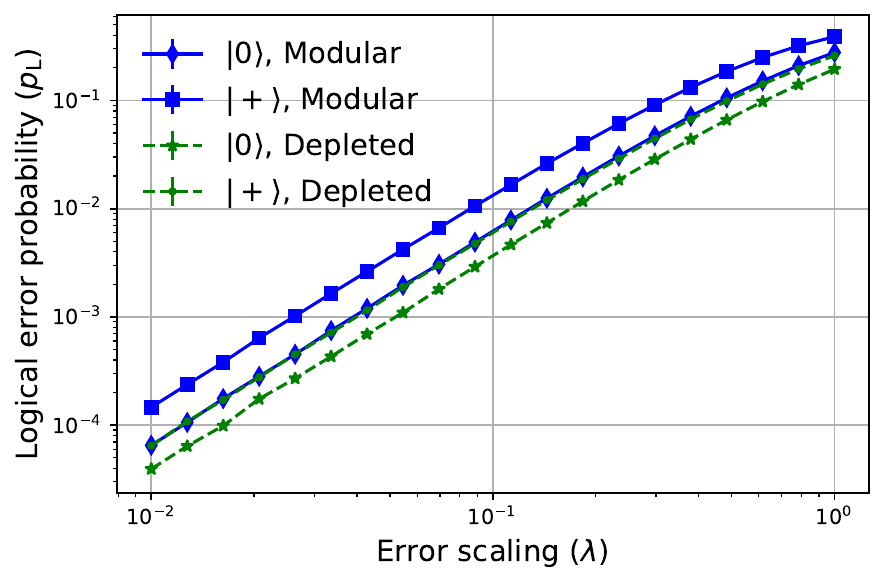}
    \caption{Comparison of the logical teleportation infidelity of \textit{fully modular} against the \textit{depleted} teleportation scheme for the surface-17 code for scaling down all physical error parameter by a factor $\lambda$. The decoding is done using MWPM. Error bars are included.}
    \label{fig:perf_Mod_vs_dep}
\end{figure}

We perform stabilizer simulations for the logical teleportation of a surface-17 code state, comparing the \textit{fully modular} and the \textit{depleted} scheme to evaluate the potential performance gain due to the reduction of the circuit depth.
Based on the extracted syndrome information, we use a minimum weight perfect matching (MWPM) decoder to determine the correction. We find that the logical teleportation is improved by a factor of $\sim$2 for going from the \textit{fully modular} to the \textit{depleted} scheme for a range of physical noise parameters, which is obtained by scaling all noise rates homogeneously by the factor $\lambda$, see~\cref{fig:perf_Mod_vs_dep}. The teleportation fidelity of the $\ket{+}_L$ state is thereby consistently worse compared to the $\ket{0}_L$ state. This imbalance can originate from the noise bias towards $Z$-type errors during the idling time in the protocol combined with the fact that for $\ket{+}_L$ state teleportation there is an increased circuit volume in which a logical error can accumulate. In particular, a logical $Z$ error of the source logical qubit also leads to a logical $Z$ error on the logical target qubit.

\subsection{In-sequence-measurement-based adaptive error correction }
\label{subsec:Adaptive_stab_meas}
A standard QEC round for the surface code, as described above, consists of two consecutive measurements of all stabilizers in the order $S^Z\rightarrow S^X\rightarrow S^Z\rightarrow S^X$. Not all of these measurements have to be performed in order to determine a correction fault-tolerantly. 
Depending on the measurement result of the first two measurement outcomes, the latter two stabilizers measurements of $Z$ and $X$ type can potentially be omitted. They only have to be executed if the earlier stabilizer measurement of respective $Z$ or $X$ type differ from the expected stabilizer value. The expected stabilizer value can be the measurement outcome of the preceding QEC round or of the value given by the state preparation.
As a consequence, after determining the outcome of the first measurement of $S^Z$ and $S^X$, on a logical state one can determine whether a second $S^Z$ and $S^X$ measurement is necessary.

This potential reduction of the number of QEC rounds can be understood from a FT point of view. Given the projective initialization of all stabilizers, an initial set of stabilizer values, the Pauli-frame, is fixed.
In general, errors are indicated by an alteration of the syndrome. However, a single syndrome readout is not always sufficient to determine a correction in a FT way because of the possibility of faults in the middle of the stabilizer measurement scheme that can lead to a corrupted syndrome combined with data qubit errors. Consequently, a committed correction based on the corrupted syndrome decision could lead to logical failure.
Following this reasoning, the respective stabilizer species on the respective code patch must be remeasured to confirm the previous outcome. Naturally, in such a second round, faults can occur as well. Still, from a FT perspective, the outcome of the repeated round can be trusted as at least one fault must already have happened during the first measurement round or before, and a FT protocol for a distance $d=3$ code only needs to ensure that every single fault remains correctable. Note that strictly speaking not for every alteration of the syndrome an additional round would be required to correct all single faults. We however do not distinguish these cases and repeat upon every alteration as this enables to apply this scheme easily for larger distances as well. A more careful selection of syndromes which require a remeasuring would leave the qualitative picture intact on when the adaptive logic can be of use.

To summarize, if the first stabilizer measurement agrees with the expected outcome, the second measurement can be omitted. Consequently, the number of gate operations and also execution time of the whole protocol can be reduced on average upon deciding adaptively and in-sequence on whether another stabilizer measurement is performed or not. Which stabilizer measurement rounds are effected in detail can be seen in~\cref{fig:combined_scheme}~(a) and~\cref{fig:combined_scheme}~(b), indicated by the gray boxes.
This alteration of the circuitry requires the experimental capability of deciding which gate sequence to be run next based on the current readout data.
A complementary work has shown a similar reduction of the average circuit depth based on estimating a confidence metric that triggers the execution of additional rounds of syndrome measurement that is calculated from soft-information~\cite{akahoshi2510runtime}
\\

In the following, we investigate the adaptive variant of the \textit{depleted} teleportation scheme presented in~\cref{subsec:FT_modularity} and show in which parameter regime in-sequence decision-logic can offer an additional performance increase.
In particular, we consider the trade-off between the reduction of the average gate count and the additional overhead that is introduced due to the in-sequence decision-logic. Such an overhead is to be expected, as the in-sequence decision-logic will incur latency, i.e. time until the conditional gate execution can continue. During this time of signal processing and information transmission, the respective qubits will idle and experience dephasing and relaxation. We assume a reference latency time of $1\upmu$s and scale it by a dimensionless parameter $\eta$. Latency values corresponding to $\eta\lessapprox 1$ are in a reasonable regime considering modern superconducting qubit quantum processors including their control architecture,  optimized for real-time feedback~\cite{gebauer_state_2020,Caune2026,song_constant-depth_2025_L,steffen_deterministic_2013}.
In addition to the idling during fixed time periods that are not affected by the adaptive scheme, due to the adaptive scheduling of syndrome extraction rounds the code patches may additionally need to idle while stabilizers in the other code patch or the ancilla qubits of the same patch are being measured.

The reduction of the average gate count, on the other hand, decreases the overall logical error rate. To leading order, the average gate count decreases approximately linearly with the gate error rates and idling error rates, for the expected noise model (see Appendix~\ref{sec:noise_modeling}).
The intuition is that single faults can trigger additional stabilizer measurements. Consequently, when all circuit error rates are scaled by a common factor $\lambda$, the corresponding reduction in the average gate count is expected to scale linearly with $\lambda$, provided the error rates remain in a regime where higher-order correlated errors are negligible. The uniform scaling of all circuit error parameters by a factor $\lambda$ serves as a simplified model for projected improvements in superconducting qubit platforms~\cite{Acharya2024,abughanem_ibm_2025}.

\begin{figure}
    \centering
    \includegraphics[width=1\columnwidth]{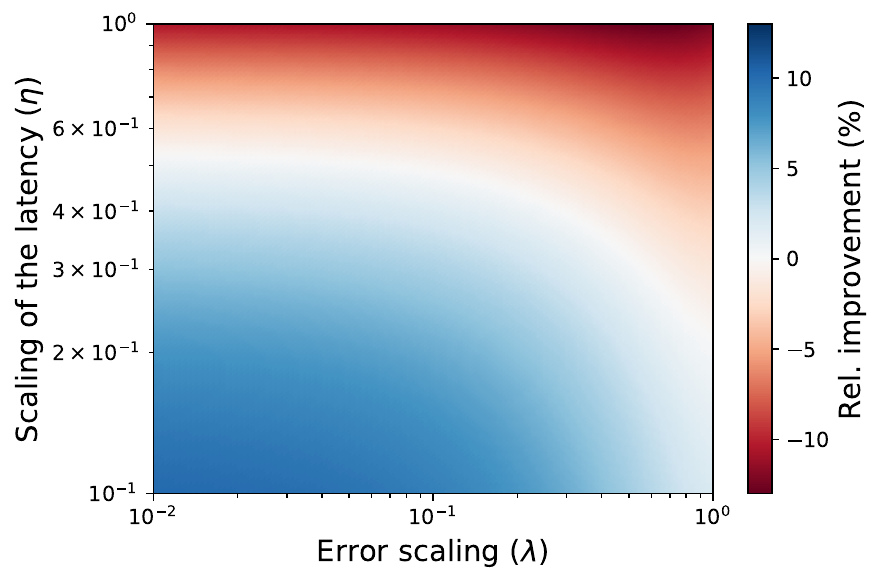}
    \caption{Relative advantage in terms of logical teleportation error rate for comparing the adaptive with the non-adaptive \textit{depleted} surgery scheme. All error parameters are scaled by $\lambda$ with respect to the reference values, and the latency of the decision logic of the adaptive scheme is scaled by $\eta$ with respect to a reference value of $1~\upmu$s. An interpolation is shown that is calculated from a grid of 20$\times$20 points.}
    \label{fig:adaptive_scan}
\end{figure}

We consider the relative difference in the logical error rate between the adaptive and non-adaptive \textit{depleted} scheme to map out the regime for which the adaptive stabilizer measurements are expected to be advantageous. The simulation results of the experimentally motivated noise model for teleporting $\ket{0}_L$ are shown in~\cref{fig:adaptive_scan}. We find that for near-term realistic improvements of the physical noise sources, $\lambda\in[0.1,1]$, the decision latency should not be larger than $\sim200\,$ns for the adaptive scheme to outperform the standard one. In the regime of lower physical error rates $\lambda\in[0.01,0.1]$, this latency requirement is relaxed to $\sim500\,$ns where the required latency to achieve an advantage levels off. We find that for low error rates and vanishing latencies, the maximal improvement of the adaptive over the standard scheme approaches $\sim10\%$. This saturation can be understood by considering the low-$\lambda$ scaling of the logical error rate which is given by $p_L^{\text{non-adapt}}=c^{\text{non-adapt}}\,\lambda^2+\mathcal{O}(\lambda^3)$ for the non-adaptive depleted teleportation protocol. For the adaptive scheme, the leading-order prefactor becomes dependent on the error rate $c^{\text{adapt}}\equiv c^{\text{adapt}}(\lambda)$ due to the varying number of average fault locations. Therefore, for sufficiently low error rates $c^{\text{adapt}}(\lambda)<c^{\text{non-adapt}}$. In the limit $\lambda\rightarrow 0$, only one round of stabilizer measurements is executed, hence $\lim_{\lambda\rightarrow 0} c^{\text{adapt}}(\lambda)=\Tilde{c}^{\,\text{adapt}}$. Therefore, the maximal improvement of $\sim10\%$ is given by 
\begin{equation}
    2\,\frac{c^{\text{non-adapt}}-\Tilde{c}^{\,\text{adapt}}}{c^{\text{non-adapt}}+\Tilde{c}^{\,\text{adapt}}},
\end{equation}
which is reached asymptotically for $\lambda\rightarrow 0$ and for $\eta=0$.

An interesting observation is that for large error rates, the maximal latency that can be tolerated to still achieve an advantage decreases strongly. This is despite the fact that every circuit element that is not to be conducted due to the adaptive scheme has a larger noise level.

For larger values $\lambda > 1$, this linear relation breaks down, as additional stabilizer measurements are effectively always triggered, removing the advantage of adaptivity. In this regime, the adaptive scheme is expected to become disadvantageous compared to the standard one, since it predominantly introduces additional idling errors during the decision latency, thereby increasing the overall logical error rate. 
In contrast, in the limit $\lambda \ll 1$, the benefit of the adaptive scheme saturates, since additional stabilizer measurements are rarely required.

\section{Higher distance, higher width projection}\label{sec:high_distance}
As logical error rates must be suppressed beyond what can realistically be achieved by reducing physical error rates alone at low code distances, it becomes necessary to scale quantum error-correcting codes to higher distances~\cite{eisert_mind_2025}. In this regime, the logical error rate of primitive operations such as lattice surgery becomes a key factor. In this section, we therefore explore the scalability of the \textit{modular} state teleportation protocol. This concerns the scaling of the code distance $d$ of the two logical surface code patches, for which the logical error rate is expected to decrease as $p_L \sim (p/p_{\text{th}})^{(d+1)/2}$ for physical error rates below threshold, $p < p_{\text{th}}$.
For this extension of the protocol, the code distance $d$ scales together with the number of stabilizer measurement rounds during the merging operation in lattice surgery. The surface code patches are thereby separated by a linear array of $d$ additional data qubits which are used in the merging operation. We denote such a merging region as having a width $w = 1$, see~\cref{fig:scaling_up_layout}.
For logical algorithms on 2D planar qubit architectures implemented via lattice surgery, the routing of logical qubits becomes relevant~\cite{Thomsen2024,litinski_game_2019,Marton2025}. In this setting, surface code patches that need to interact may not be adjacent, but instead can be separated by regions of width $w$ data qubits, as depicted in~\cref{fig:scaling_up_layout}.
We investigate how the logical error rate depends on the separation width $w$ and find a clear increment of the logical error rate with $w$, see Fig.~\ref{fig:d_w_projection}.
Hence, we conclude that a single column of data qubits between the surface code patches remains optimal, as increasing the width of the merging region leads to a higher logical error rate; more details are given in Sec.~\ref{subsec:width}.

We note that some of us present further studies on thresholds and in-homogeneous error rates for varying $w$ in Ref.~\cite{Marton2025}, based on a statistical physics based scaling theory. 
Here, in contrast, our focus lies on providing estimates for an experimentally realistic noise model rather than exploring fundamental dependencies.
Furthermore, we find that achieving exponential suppression of the logical error rate for teleporting both $\ket{0}_L$ and $\ket{+}_L$ requires improving all physical error parameters at least by an error scaling factor of $\lambda_{\rm th} \approx 0.55$, where $\lambda=1$ corresponds to the experimentally observed noise for the distance $d=3$ surface code memory experiment from Ref.~\cite{krinner_realizing_2022}, see also~\cref{sec:noise_modeling}. This threshold can be read off from the large distance simulations presented in~\cref{fig:threshold} for $\ket{0}_L$ and $\ket{+}_L$ in panels (a) and (b), respectively; more details can be found in Sec.~\ref{subsec:large_d}.
\subsection{Scaling the size of the merging region}\label{subsec:width}
\begin{figure}[t]
    \centering
\includegraphics[width=0.45\textwidth]{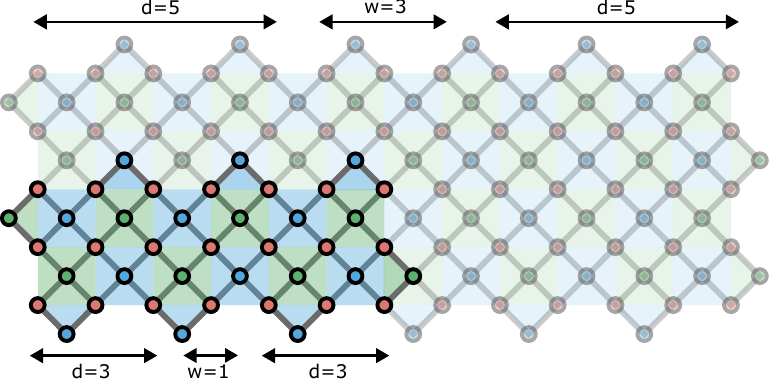}
    \caption{The layouts of rectangular chips with different parameters. The region of the chip shown in saturated colors contains two distance three surface code patches ($d=3$) separated by a single column of data qubits ($w=1$). The entire shown chip contains two distance five surface codes ($d=5$) separated by three columns of data qubits ($w=3$).}
    \label{fig:scaling_up_layout}
\end{figure}

In the following, we will show that increasing the number of physical qubits and the correspondingly associated stabilizers between the two surface code patches does not improve the logical performance of the lattice surgery. We find, similar as in~\cref{subsec:FT_modularity} for the considerations on the temporal extent of the surgery (number of stabilizer measurements during the merging), that it is best to use the smallest amount of merging qubits as long as the fault distance is preserved. For lattice surgery between two distance-$d$ surface code patches, this implies having a single column of $d$ additional data qubits as the merging region. This corresponds to the lattice surgery layout, which we consider for the $d=3$ surface-17 code in Sec.~\ref{sec:lattice_surgery}, see~\cref{fig:scaling_up_layout}.

A priori, it might not be obvious that the best strategy is to resort to the minimal width $w=1$ at arbitrary distance. In principle, one might expect that having more data qubits in the merging region may enhance the protection of the parity measurements. However, as the gates and qubits are noisy, increasing the width of the interface region also increases the effective noise strength that the whole protocol experiences.
To answer this question of the optimal width quantitatively, we simulate the \textit{modular} teleportation scheme employing the previously used, experimentally motivated circuit level noise model described in~\cref{sec:noise_modeling}.

\begin{figure}
    \centering
    \includegraphics[width = .5\textwidth]{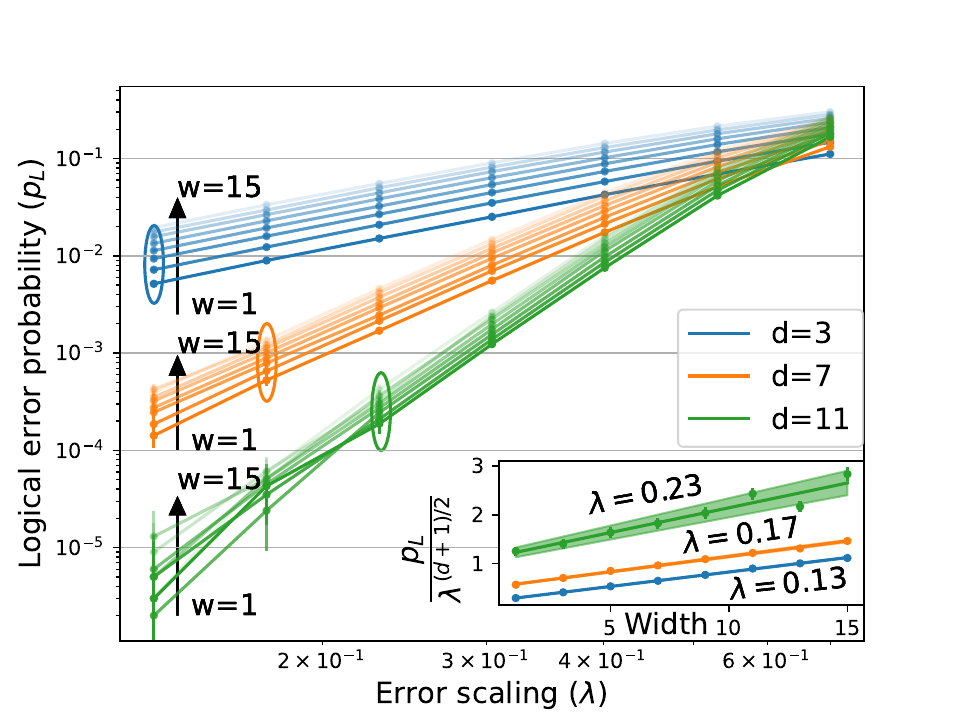}
    \caption{Logical error rate of the teleportation of the logical $\ket{0}_L$ state, using the modular scheme for $d=3,7,11$ and $w=1,3,...,15$. The logical error rate always increases as the width $w$ grows for any of the code distances observed. In the inset the $w$ dependence of the rescaled logical error rate is shown for three \(\lambda\) values (encircled points). Points are indicating simulation data for \(\lambda=0.13,~d=3\) (blue), \(\lambda=0.17,~d=7\) (orange), \(\lambda=0.23,~d=11\) (green), while continuous lines are linear fits in the form of Eq.~\eqref{eq:first_order_fit}.}
    \label{fig:d_w_projection}
\end{figure}

The setting is illustrated in~\cref{fig:scaling_up_layout}. We consider a rectangular chip with width \(2d+w\) and height \(d\), and a total of \(d(2d+w)-1\) ancilla qubits. We scale up the \textit{modular} teleportation scheme as follows: (i) We perform the initialization step by means of a single round of stabilizer measurements. (ii) During the merging step, we perform $d$ rounds of $X$ and $Z$ stabilizer measurements. (iii) After the split, we immediately measure the logical operators, as additional rounds of stabilizer measurements do not increase the fault-distance.
In Fig.~\ref{fig:d_w_projection}, we show simulations of the teleportation of the logical $\ket{0}_L$ state for code distances \(d=3,7,11\) and widths \(w=1,3,5,\ldots,15\),
in the range of \(\lambda=0.13 - 0.7\) to show the quantitative sub-threshold behavior. We note that increasing the size of the interface region always increases the logical error of the teleportation protocol, regardless of the code distance. In fact, there is a linear dependence between the logical error rate and the width \(w\), as it is captured by the lowest order expansion of the logical error rate~\cite{chamberland_universal_2022}
\begin{align} \label{eq:first_order_fit}
    p_L = \big(A(d)w + B(d)\big)\lambda^{(d+1)/2} + \ldots,
\end{align}
where \(A(d)\) and \(B(d)\) are the combinations of combinatorial factors and $d$-dependent powers of error rates.
To verify this linear scaling, we fit linear curves to the rescaled data, $p_L/\lambda^{(d+1)/2}$,  for the smallest \(\lambda\) values where the statistical uncertainty is not too high. The linear fits are shown in the inset of Fig.~\ref{fig:d_w_projection} for code distances $d=3,7,11$ and the corresponding \(\lambda\) values \(\lambda = 0.13,0.17,0.23\).

\subsection{Expected logical performance of larger distance}
\label{subsec:large_d}

In Sec.~\ref{subsec:FT_modularity} we have investigated the performance of the teleportation of the logical $\ket{0}_L$ and $\ket{+}_L$ states with the \textit{modular} scheme for $d=3$. Here we now use this protocol and scale up the code distance $d$, while keeping the width of the region of qubits used for the merging process at the optimal value, $w=1$. We use experimentally motivated noise parameters as in the previous sections (see~\cref{sec:noise_modeling} for details) and we omit the $d-1$ rounds of stabilizer measurement before and after the merging operation in the \textit{modular} scheme, just as we have done in~\cref{subsec:width}.

\begin{figure}
    \centering
    \includegraphics[width=0.5\textwidth]{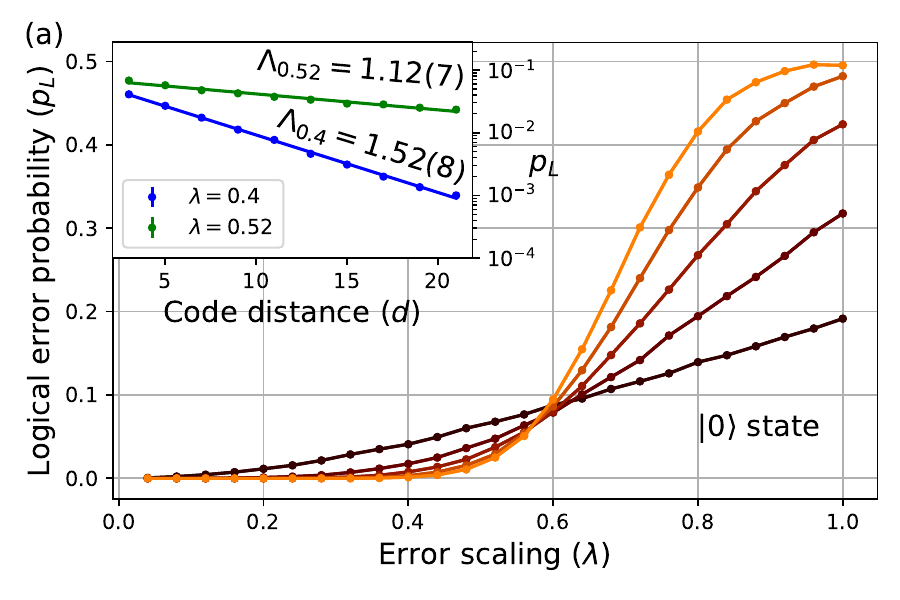}
    \includegraphics[width=0.5\textwidth]{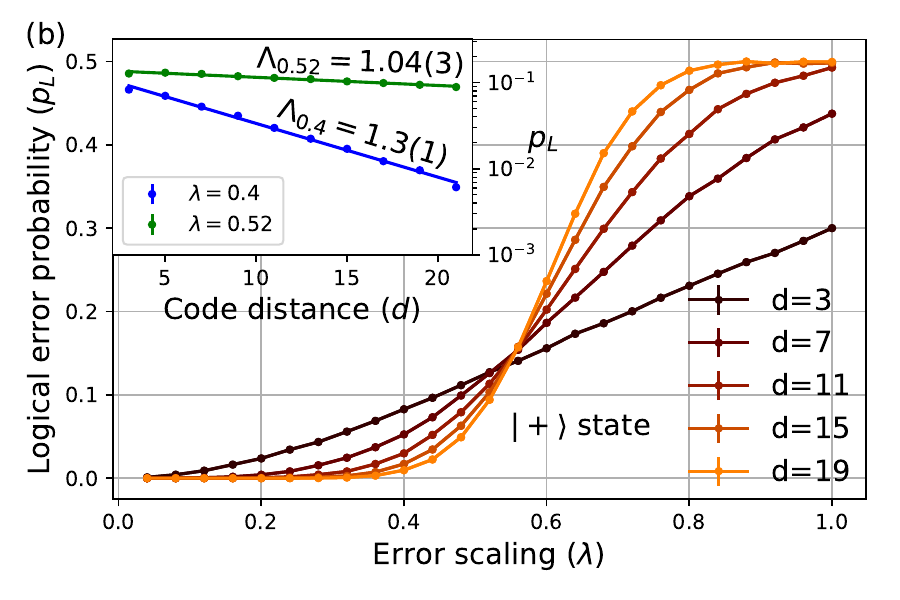}
    \caption{The logical error probability of the teleportation of the logical $\ket{0}_L$ and  $\ket{+}_L$ state in panels (a) and (b) respectively, by using the modular scheme for physical error parameters scaled down by a factor $\lambda$. For error scaling-factors lower than $\lambda_{th} \approx 0.55$ the logical error rate can be suppressed exponentially by scaling up the fault-distance of the protocol, as is shown in the inset for $\lambda = 0.4$ and $\lambda = 0.52$.}
    \label{fig:threshold}
\end{figure}

Our numerical results for code distances from $d=3$ up to $d=19$ indicate that to reach the error correction threshold the physical noise parameters have to be reduced by at least a factor of $\lambda_{\rm th} \approx 0.55$, relative to the parameters in Ref.~\cite{krinner_realizing_2022}. This means that if the physical noise is below this threshold, the logical error probability can be suppressed exponentially for both logical basis states by increasing the distance. To quantify the exponential suppression, we calculate the error suppression factor $\Lambda_{\lambda}$ \cite{google_quantum_ai_and_collaborators_quantum_2025}, given for every fixed $\lambda$ by
\begin{align}
    p_L = \Big(\dfrac{1}{\Lambda_{\lambda}}\Big)^{(d+1)/2}.
\end{align}
We determine the error suppression factors by fitting a linear curve to $\ln(p_L)$ as a function of $d$. Our numerical results and the error suppression factors for $\lambda =0.4$ and $\lambda = 0.52$ are shown in Fig.~\ref{fig:threshold}. These scaling factors confirm that $\lambda = 0.52$ is just barely below the effective threshold for this protocol and noise model, offering only a mild noise suppression of $~4\%$ and $~12\%$ for the $\ket{+}_L$ and $\ket{0}_L$ state respectively per step of increasing the distance by 2, i.e. $d\rightarrow d+2$. At the same time, a stronger reduction of the reference error rate by $60\%$ achieves a logical noise suppression by about $30\%$ and $50\%$ for the $\ket{+}_L$ and $\ket{0}_L$ state respectively, when increasing the code distance by 2.
\section{Conclusions and Outlook}

In this work, we have conducted a theoretical study of surface code-based logical teleportation protocols based on lattice surgery, in which a single logical state is teleported using entanglement with a second logical qubit. We have described how to execute the building blocks involved in the lattice surgery protocol, namely logical qubit initialization, merging, splitting, and logical measurement. As a concrete case study, we have analyzed the anticipated performance of the protocol for a superconducting qubit based architecture by conducting numerical simulations for an experimentally informed multi-parameter Pauli error model.
As potential variants of the protocol, we have considered adaptive stabilizer measurements, where the execution of a stabilizer measurement is conditioned on previous measurement outcomes. Here, we have analyzed the trade-off between latency times and error rates, identifying the regime in which the adaptive scheme is advantageous.
Furthermore, we have explored the potential benefits of increasing the size of the interface region between the two surface code patches. Our simulations show that the teleportation success probability does not increase when the size of the interface region is increased in any of the scenarios we consider. 

We present this work as a projection for near-term experiments with superconducting qubits, with a focus on the comparative analysis of different lattice surgery variants and their fault-tolerance aspects, rather than on describing a specific setup with a sophisticated microscopic error model. Extensions of the present work could, however, include accounting for leakage errors, and the incorporation of leakage reduction units~\cite{, Miao2023,lacroix_fast_2025} in the circuits. These gadgets allow one to prevent leakage errors from spreading and to ensure that leaked qubits are returned into the computational subspace, and they in turn require the correction and decoding routines have to be adapted correspondingly~\cite{varbanov_leakage_2020}. Furthermore, it is possible to analyze the influence of cross-talk errors~\cite{Zhou2025,sarovar_detecting_2020} and correlated noise on the expected logical performance. Exploring the potential benefit of adaptive over deterministic protocols in more complex surgery-based protocols constitutes a further potential extension of the present work.

\section{Acknowledgement}
We thank all members of the SuperMOOSE consortium under the ELQ-call for useful and stimulating discussions.
All authors gratefully acknowledge support by the Intelligence Advanced Research Projects Activity (IARPA) and the Army Research Office, under the Entangled Logical Qubits program through Cooperative Agreement Number W911NF-23-2-0212. 
LB, AM, LC and MM also acknowledge the support by the Deutsche Forschungsgemeinschaft (DFG, German Research Foundation) under Germany’s Excellence Strategy ‘Cluster of Excellence Matter and Light for Quantum Computing (ML4Q) EXC 2004/1’ 390534769 as well as under the Schwerpunktprogramm 2514, project number 541030623, and support by the Federal Ministry of Research, Technology
and Space of Germany (BMFTR) through the project SQale. This research is also part
of the Munich Quantum Valley---Hardware Adapted Theory, which is sup-
ported by the Bavarian state government with funds
from the Hightech Agenda Bayern Plus.
The authors gratefully acknowledge the computing time provided to
them at the NHR Center NHR4CES at RWTH Aachen
University (project number p0020074).
The views and conclusions contained in
this document are those of the authors and should not
be interpreted as representing the official policies, either
expressed or implied, of IARPA, the Army Research Office, or the U.S. Government. The U.S. Government is
authorized to reproduce and distribute reprints for Government purposes notwithstanding any copyright notation
herein.

\appendix
\section{Circuits for logical teleportation of $d=3$ surface code}\label{sec:circuits}
\begin{figure*}[t]
    \centering
    \includegraphics[width=0.9\textwidth]{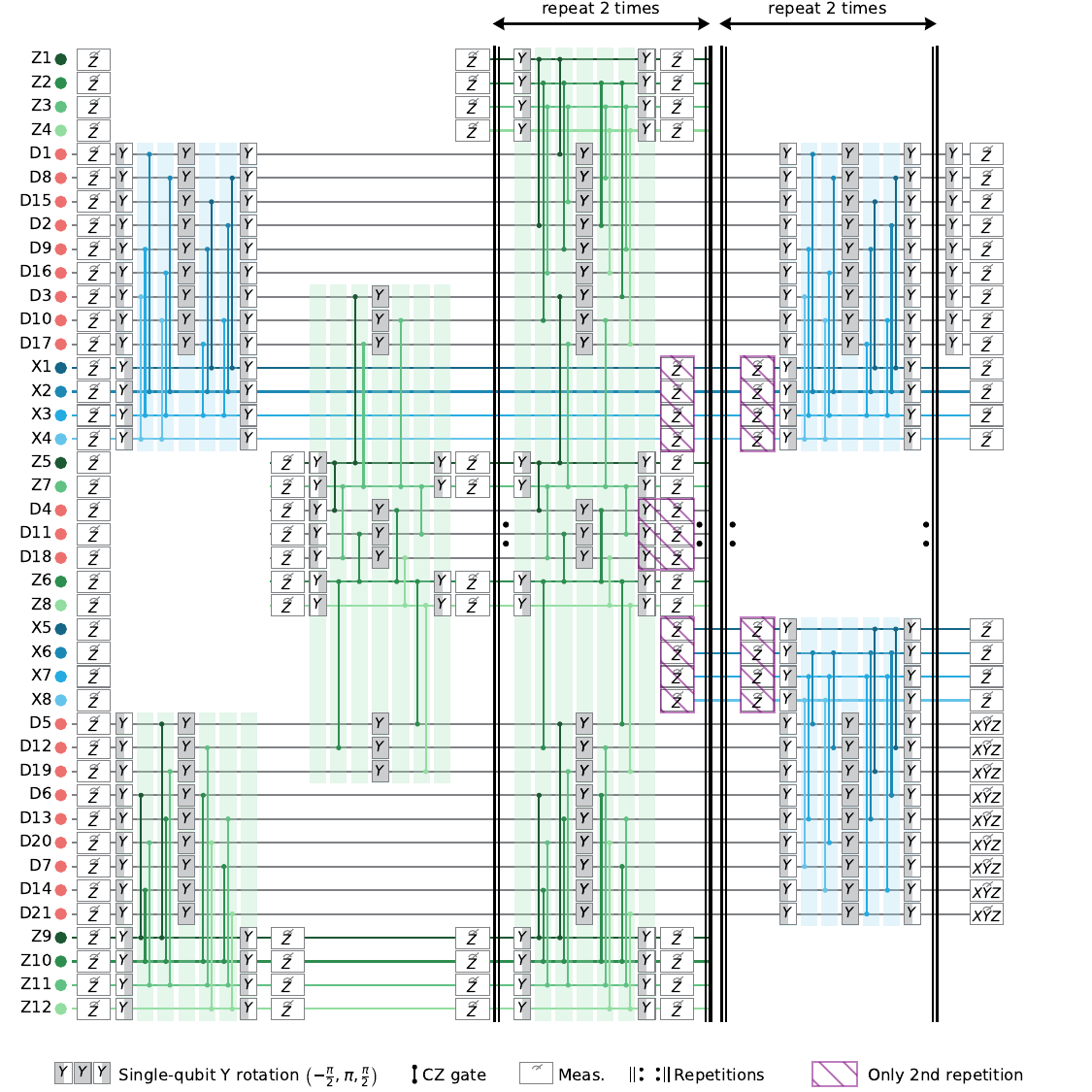}
    \caption{Full circuit of the \textit{depleted} logical teleportation protocol (see~\cref{fig:combined_scheme}~(b)) in the surface-41 chip.}
    \label{fig:circuit_depleted}
\end{figure*}

In this section, we provide the precise circuitry decomposed into the natively available experimental transmon gate set.
To conduct realistic simulations of these circuits for measuring the stabilizer generators of the surface code, following Ref.~\cite{krinner_realizing_2022}, we consider: (i) a single-qubit gate duration of 40~ns, (ii) a two-qubit gates with duration of 100~ns, and (iii) a readout of~400~ns. 
We show the full circuits in~\cref{fig:full_circuit_a,fig:full_circuit_b,fig:circuit_depleted}. Since the bulk of stabilizers in the rotated surface code is weight-4, the syndrome extraction circuit requires four distinct time slots for two-qubit gates per round of syndrome extraction.

\begin{figure*}[t]
    \centering
    \includegraphics[width=0.9\textwidth]{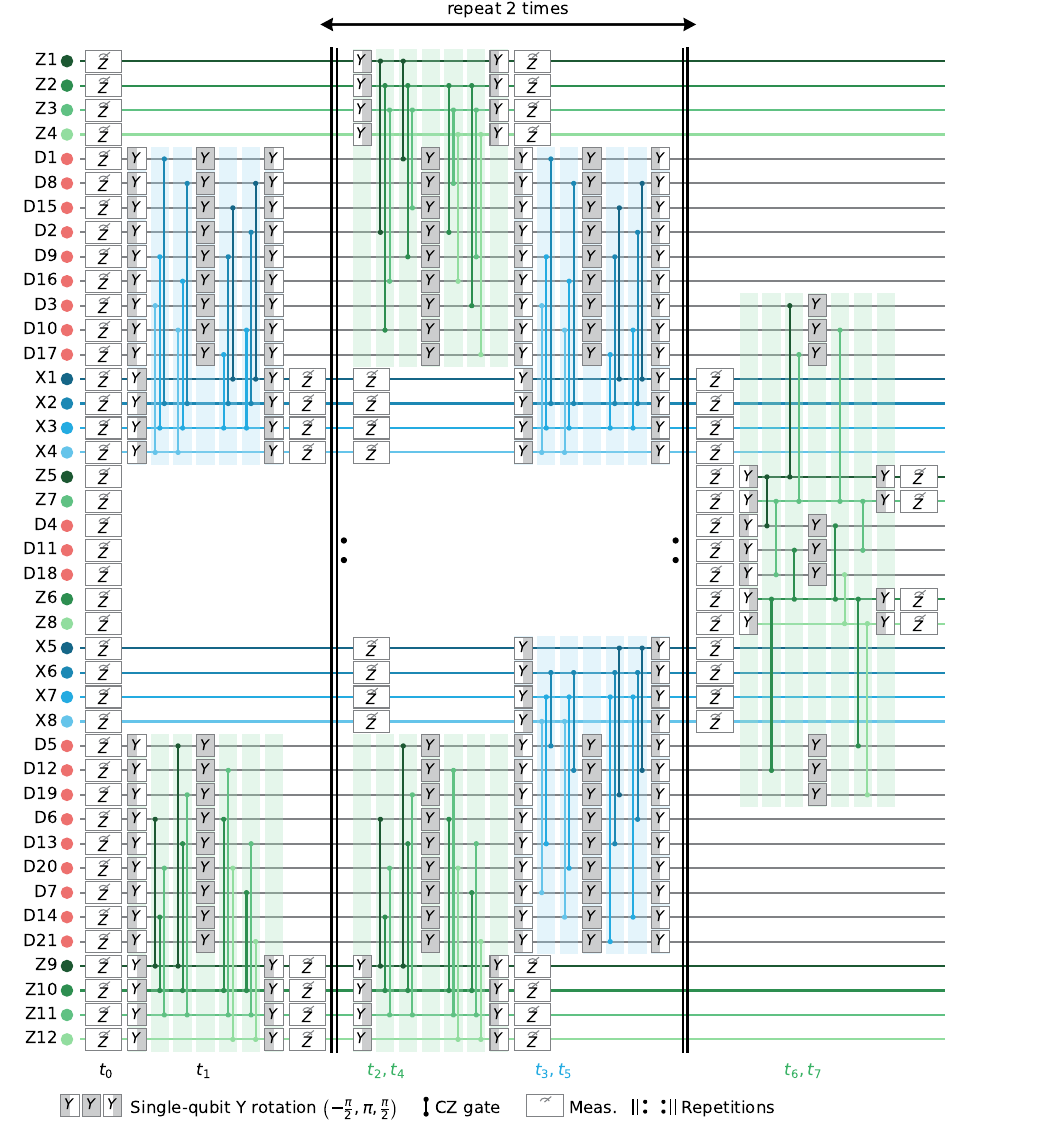}
    \caption{Initialization followed by a full QEC round and the merging step of the full logical teleportation protocol (see~\cref{fig:combined_scheme}~(a)) in the surface-41 chip layout. The qubit labels correspond to the ones shown in Fig.~\ref{fig:surface_41_layout}. The gate set is composed of single qubit $\exp(-i\alpha Y/2)$ rotations with $\alpha=(-\pi/2,\pi,\pi/2)$ and CZ gates. The circuit continues in Fig.~\ref{fig:full_circuit_b}.}
    \label{fig:full_circuit_a}
\end{figure*}

\begin{figure*}[t]
    \centering
    \includegraphics[width=0.9\textwidth]{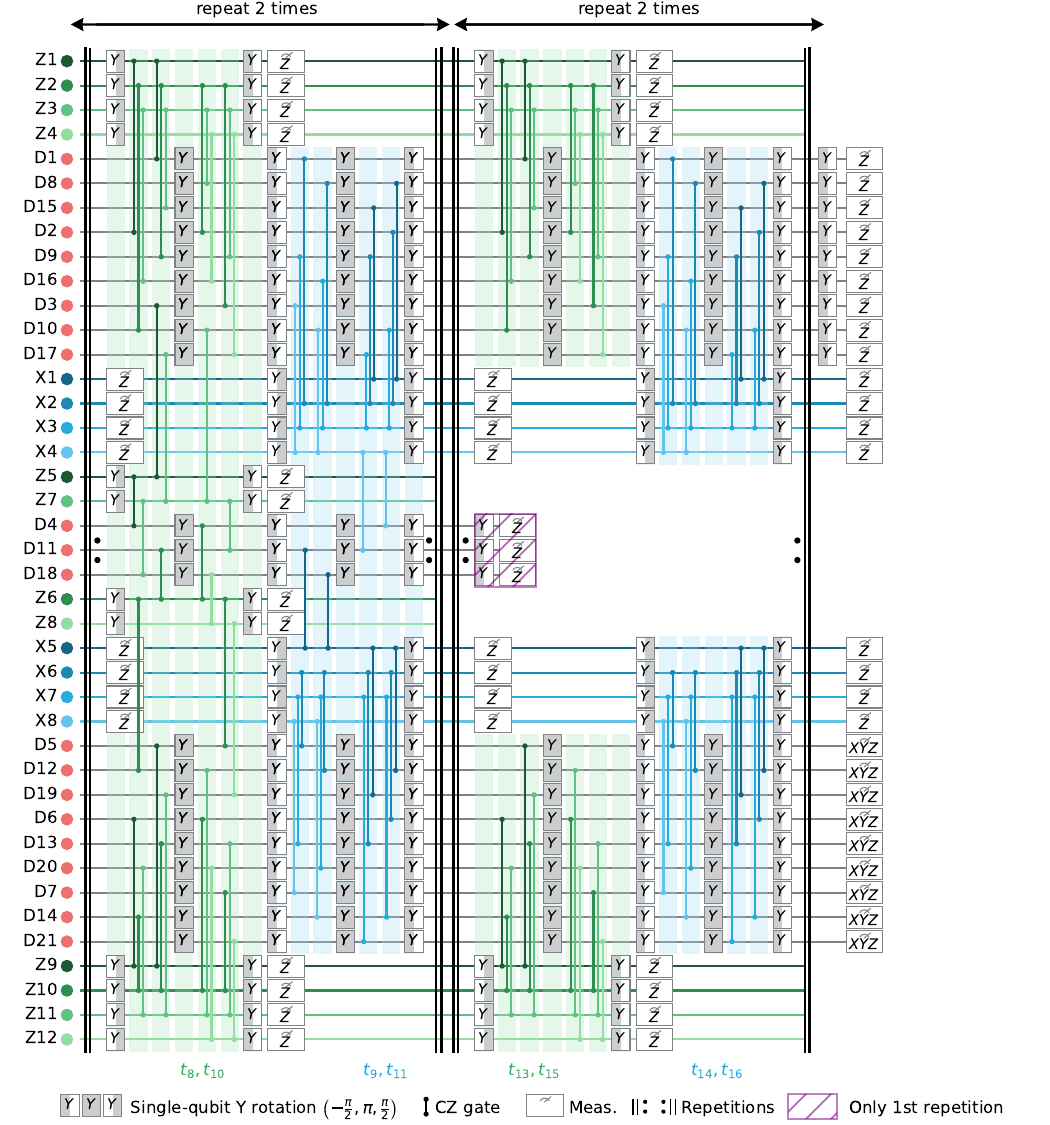}
    \caption{Continuation of the circuit from \cref{fig:full_circuit_a}. Full QEC round after merging followed by splitting step and an additional QEC round, see protocol in~\cref{fig:combined_scheme}~(a). The qubit labels correspond to the ones shown in Fig.~\ref{fig:surface_41_layout}.}
    \label{fig:full_circuit_b}
\end{figure*}

\newpage
\clearpage

\section{Noise modeling}\label{sec:noise_modeling}

\subsection{Standard stabilizer simulations}
\label{subsec:standard_noise_modeling}
In order to determine the performance of the protocols described above, we perform Clifford simulations assuming circuit level Pauli noise. The noise model thereby includes the following processes that each have fixed base error parameters, see~\cref{tab:ref_error_rates}
\begin{enumerate}
  \item Initializing the complementary qubit state with respect to the intended one, i.e. $\{\ket{0},\ket{1},\ket{+},\ket{-}\} \rightarrow \{\ket{1},\ket{0},\ket{-},\ket{+}\}$.
  \item Recording the complementary measurement outcome with respect to the actual qubit state.
  \item Applying an error drawn from $\{ X, Y, Z\}$ after a single-qubit gate with equal probability.
  \item Applying an error drawn from $\{I, X, Y, Z\}^{\otimes2}\backslash \, I\otimes I$ after a two-qubit gate with equal probability.
  \item Applying a $Z$ error after an idling time of duration $\tau$ with probability $\frac{1}{2}\left(1-e^{-\tau/T_2} \right)-\frac{1}{4}\left(1-e^{-\tau/T_1} \right)$ and an $X$ or $Y$ error with probability $\frac{1}{4}\left(1-e^{-\tau/T_1} \right)$ respectively~\cite{tomita_low-distance_2014}.
\end{enumerate}
 The reference error parameters in~\cref{tab:ref_error_rates} correspond to the average gate and measurement error rates and characteristic time scales, obtained in the quantum memory experiment~\cite{krinner_realizing_2022}, based on a repeated syndrome extraction for surface-17 implementation in a superconducting transmon architecture~\cite{krinner_realizing_2022}. 
After measuring an ancilla qubit an unconditional reset operation to $|0\rangle$ is applied, such that the recorded measurement outcome is inverted with the probability of the measurement error rate while the post-measurement qubit state is corrupted with the probability of the initialization error.
When ancilla resets are not implemented, usually one extra stabilizer measurement is needed to ensure fault-tolerance~\cite{geher2025}. Hence, in the setting where no reset operation is available, the whole teleportation protocol presented in this work would need to be modified by adding extra stabilizer measurements to obtain a fault-tolerant protocol.
 The motivation of utilizing these error parameters is to use them as a reference point, to project the performance of the different lattice surgery protocols for currently available and near-term experimental superconducting hardware. Furthermore, we are interested in scaling down of the reference error parameters homogeneously by a factor $\lambda\in (0,1]$ to get an estimation of the expected behavior of the logical error rate for future experimental improvements. Here $\lambda=1$ corresponds again to the present-day values from~\cite{krinner_realizing_2022}.

\begin{table}[]
    \centering
    \begin{tabular}{c|c}
        Initialization error rate& 0.09\%\\
        \hline
        Measurement error rate& 2.2\%\\
         \hline
         Single qubit gate error rate& 0.09\%\\
         \hline
         Two qubit gate error rate& 1.5\% \\
         \hline
         Effective coherence time $T_2$ & 47 $\upmu$s\\
        \hline
         Lifetime $T_1$ & 32.5 $\upmu$s \\
    \end{tabular}
    \caption{Set of average physical reference error rates as observed in the experiment~\cite{krinner_realizing_2022}. Here, the measurement error rate which is used in the simulation corresponds to the 3-state readout misclassification probability. It emerges from the discretization of the measurement outcome into the three classes $\{0,1,>1\}$ based on the recorded quadrature expectation value from the dispersive readout of the transmon~\cite{Blais21,Chen2023}. The effective coherence time is defined as $(T_2)^{-1}=(2T_1)^{-1}+(T_\phi)^{-1}$ where $T_\phi$ is the bare coherence time.}
    \label{tab:ref_error_rates}
\end{table}
\begin{table}[]
    \centering
    \begin{tabular}{c|c}
         Single qubit gate& 40 ns  \\
         \hline
         Two qubit gate& 96 ns\\
         \hline
         Measurement& 400 ns\\
         \hline
         Full cycle&1.008 $\upmu$s\\
         \hline
         Half cycle& 0.504 $\upmu$s\\
         \hline
         Latency (adaptive scheme)&1 $\upmu$s
    \end{tabular}
    \caption{Durations of the different circuit elements or building blocks in the different surgery protocols. From these times, the effective idling error rate derives~\cite{tomita_low-distance_2014}. }
    \label{tab:Duration_times}
\end{table}

\section{Non-scalability of the depletion of the lattice surgery}
\label{sec:scala_deplt}

\begin{figure*}
    \centering
    \includegraphics[width=0.4\linewidth]{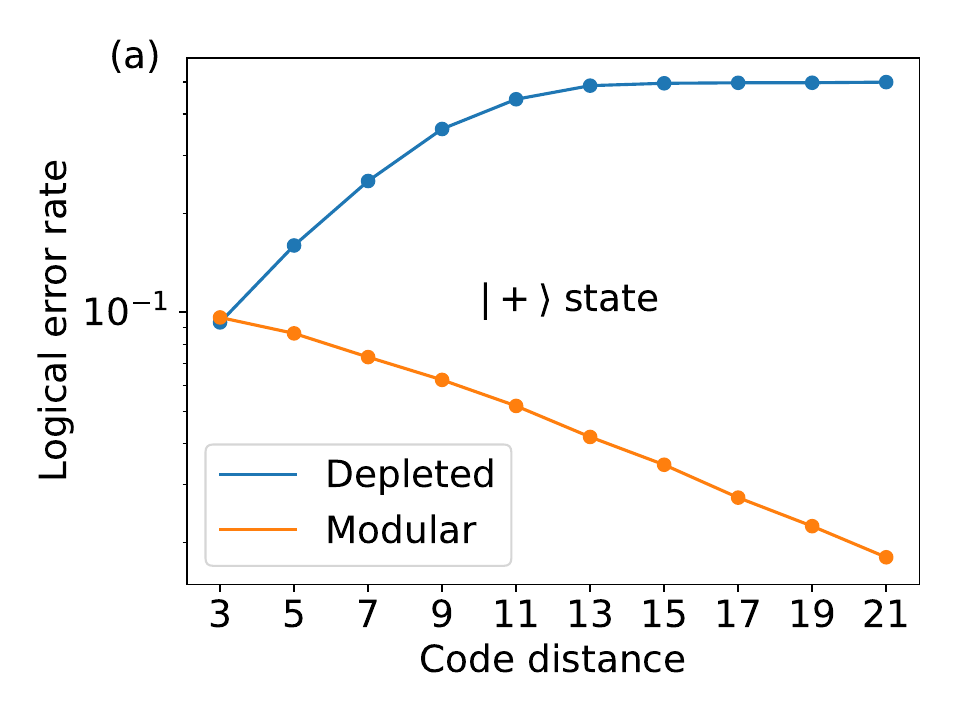}
    \includegraphics[width=0.4\linewidth]{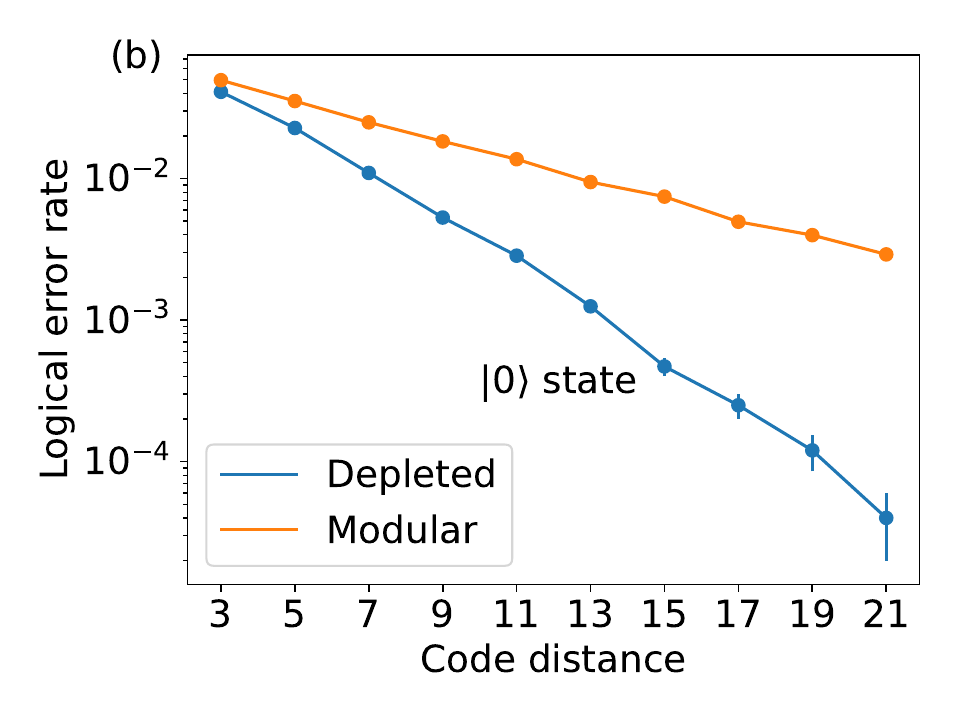}
    \caption{Comparison of the logical error rates for teleporting a $\ket{+}_L$ state in (a) and a $\ket{0}_L$ state in (b), respectively, using the depleted scheme (blue) and the modular scheme (orange) as both schemes are scaled to higher code distances. The simulations are performed for the experimentally motivated noise model with noise-scaling factor $\lambda=0.44$, below the threshold of the modular lattice-surgery protocol.}
    \label{fig:non_scalability}
\end{figure*}

As discussed in the main text, the depleted lattice-surgery scheme, in which only $d$ rounds of $Z$-type stabilizer measurements are performed during the $ZZ$ merge of two distance-$d$ surface codes, is not scalable. The reason is that, as the code distance increases, the protocol depth also increases, allowing $Z$-type errors to accumulate. This leads to a vanishing threshold for preserving the logical operator $X_L$.

We illustrate this behavior in~\cref{fig:non_scalability} by comparing the logical error rates of the scaled-up modular and depleted schemes for teleporting logical $\ket{+}_L$ and $\ket{0}_L$ states at increasing code distance $d$. For the $\ket{+}_L$ state, shown in~\cref{fig:non_scalability}(a), the depleted scheme outperforms the modular scheme only at $d=3$, but its logical error rate increases with distance. Thus, for the experimentally motivated error model considered here, the modular scheme is preferable for all system sizes with $d>3$.
In contrast, when teleporting the $\ket{0}_L$ state, only the logical observable $Z_L$ needs to be protected. As shown in~\cref{fig:non_scalability}(b), the depleted schedule consistently outperforms the modular schedule for all distances considered.

\newpage
\bibliography{references,references_2}
\end{document}